\documentclass{PoS}

\title{Full NLO electroweak corrections to Z-boson pair production at the Large 
Hadron Collider}

\ShortTitle{Electroweak corrections to Z-boson pair production}

\author{\speaker{Benedikt Biedermann}\\%\thanks{A footnote may follow.}\\
Universit\"at W\"urzburg, Institut f\"ur Theoretische Physik und Astrophysik, 
Emil-Hilb-Weg 22,
97074 W\"urzburg, Germany
\\
E-mail: \email{benedikt.biedermmann@physik.uni-wuerzburg.de}}

\abstract{
We report on a recent calculation of the full next-to-leading-order electroweak corrections
to Z-boson pair production with subsequent decays into four charged leptons. Using the complete matrix
elements at leading order and next-to-leading order in the electroweak coupling for the processes $\Pp\Pp\to\mu^+\mu^-\Pe^+\Pe^-$ and $\Pp\Pp\to\mu^+\mu^-\mu^+\mu^-$, this includes all off-shell effects of intermediate massive vector bosons and photons.
We employ a gauge-invariant splitting for the electroweak
corrections into purely weak and photonic corrections. The latter show the well-known radiative tails near kinematical thresholds or resonances. The former are generically at
the level of $\sim-5\%$ for the fiducial cross section and reach several $-10\%$ in the high-energy tails of distributions due to logarithms of electroweak origin. 
The impact of interference effects due to equal-flavour leptons in the final state can reach the order of $5\%$ in off-shell-sensitive regions.
Photon-induced contributions are included in our calculation, but turn out
to be phenomenologically unimportant.
}

\FullConference{XXV International Workshop on Deep-Inelastic Scattering and 
Related Subjects\\
		3-7 April 2017\\
		University of Birmingham, UK}

\usepackage{amsmath}
		
\newcommand{\GeV}{\unskip\,\mathrm{GeV}}

\newcommand{\TeV}{\unskip\,\mathrm{TeV}}

\def\mathswitch#1{\relax\ifmmode#1\else$#1$\fi}
\def\mathswitchr#1{\relax\ifmmode{\mathrm{#1}}\else$\mathrm{#1}$\fi}
\def\mathswitchit#1{\relax\ifmmode{#1}\else$#1$\fi}

\newcommand{\Pu}{\mathswitchr u}
\newcommand{\Pd}{\mathswitchr d}
\newcommand{\Ps}{\mathswitchr s}
\newcommand{\Pc}{\mathswitchr c}

\newcommand{\Pp}{\mathswitchr p}

\newcommand{\Pe}{\mathswitchr e}

\newcommand{\PZ}{\mathswitchr Z}

\newcommand{\recola}{{\sc Recola}}
\newcommand{\collier}{{\sc Collier}}

\newcommand{\Pb}{\mathswitchr b}

\newcommand{\ppmmee}{\Pp \Pp \to \mu^+\mu^-\Pe^+\Pe^- +X}

\newcommand{\ppmmmm}{\Pp \Pp \to \mu^+\mu^-\mu^+\mu^- +X}

\newcommand{\mr}{\mathrm}

\begin{document}

\section{Introduction}

Vector-boson pair production represents an important class of processes 
at the Large Hadron Collider (LHC). Due to their sensitivity to the triple
gauge-boson couplings these processes are sensitive to effects from physics beyond the
standard model. Moreover, they constitute an important background to the decay of the
Higgs-boson into weak gauge-boson pairs.
Most of the theoretical predictions for this process class focused so far on QCD corrections which are 
meanwhile known at next-to-next-to-leading order (NNLO) accuracy in the strong coupling \cite{Grazzini:2015hta,Grazzini:2016swo,Grazzini:2016ctr}. The next-to-leading order (NLO) electroweak (EW) corrections to vector-boson 
pair production are known in the literature for on-shell W and Z bosons 
\cite{Bierweiler:2013dja,Baglio:2013toa}. For ZZ and WW 
production, the full NLO EW corrections that fully take into account all off-shell effects have been presented recently
\cite{Biedermann:2016guo,Biedermann:2016yvs,Biedermann:2016lvg,Kallweit:2017khh}.
In this proceedings contribution, we follow Ref.~\cite{Biedermann:2016lvg} and focus on the full EW corrections to the 
production of Z-boson pairs 
with subsequent decays to four charged
leptons. Using complete matrix elements for the hadronic processes $\Pp\Pp\to\mu^+\mu^-\Pe^+\Pe^-$ and $\Pp\Pp\to\mu^+\mu^-\mu^+\mu^-$, this includes all off-shell effects of intermediate massive vector bosons and photons.

\section{Numerical setup}
The leading-order (LO) contribution constitutes of the 
quark-antiquark 
annihilation channels $\bar q q/q\bar q\to\mu^+\mu^-\Pe^+\Pe^-,\,\mu^+\mu^-\mu^+\mu^-$ at order $O(\alpha^4)$ %according to
%\begin{align}%
%
%\end{align}
with massless quark flavours $q=\{\Pu,\Pd,\Pc,\Ps,\Pb\}$.
The NLO EW corrections at order $O(\alpha^5)$ comprise virtual and real 
contributions of the
$\bar q q$ channels,  $\bar q  q/q \bar q \to \mu^+\mu^-\Pe^+\Pe^- \,(+\gamma),\;\mu^+\mu^-\mu^+\mu^- \,(+\gamma)$,
and the real photon-induced contributions with one (anti)quark and one
photon in the initial state. The latter, generically referred to as $q\gamma$ channels in the following, are obtained from the $\bar q q$ channel by crossing the real final-state photon with an initial-state (anti)quark.
The one-loop corrections are further split in a gauge-invariant way into a 
purely 
weak part and a photonic contribution. The photonic part is defined as the collection of 
all diagrams with at least one photon in the loop coupling to the external 
fermion lines. The purely weak part consists of the remaining one-loop diagrams. We remark that the IR divergences in the photonic part of the 
virtual corrections are entirely cancelled by the corresponding divergences of 
the real corrections that we isolate via the dipole subtraction formalism 
\cite{Catani:1996vz,Dittmaier:1999mb}. The purely weak NLO corrections are thus by 
construction 
IR finite with LO kinematics. We have also computed the purely photon-induced 
contribution $\gamma\gamma \to\mu^+\mu^-\Pe^+\Pe^-,\,
\mu^+\mu^-\mu^+\mu^-$. As this channel turns out to be strongly suppressed with 
respect to the LO cross section, we do not compute any corrections to this 
contribution. 

The complex-mass scheme \cite{Denner:2005fg} has been used for consistently treating the massive gauge-boson resonances.
The matrix elements have been computed with \recola~\cite{Actis:2016mpe} in combination with the tensor-integral library \collier~\cite{Denner:2016kdg}, and 
cross checked against amplitudes from the private {\sc Mathematica}
package {\sc Pole}~\cite{Accomando:2005ra} and against a private implementation based on diagrammatic methods like those
developed for four-fermion production in electron--positron collisions~\cite{Denner:2005fg}.

In the analysis presented below, we consider the LHC at a 
centre-of-mass (CM) energy of $13\TeV$. The employed input parameters
(masses and widths of the particles and coupling constants) can 
be found in Ref. \cite{Biedermann:2016lvg}. 
As parton distribution function, we use the NNPDF-2.3 NLO set including QED corrections with
$\alpha_{\rm s}(M_\PZ)=0.118$ 
\cite{Ball:2013hta}.
We consider a minimal set of selection cuts restricting
the charged leptons $\ell_i$ in transverse momentum $p_{{\rm 
T},\ell_i}$, 
rapidity $y_{\ell_i}$ and lepton separation $\Delta R_{\ell_i,\ell_j}$ according to
\begin{align}
 p_{{\rm T},\ell_i}>p_{\rm T,min}=15\GeV,\qquad |y_{\ell_i}|&<2.5,\qquad\Delta R_{\ell_i,\ell_j} = \sqrt{(y_{i}-y_{j})^2+(\Delta\phi_{ij})^2}>0.2.
 \label{eq:inc-cuts1}
\end{align}
%
%Any pair of charged leptons $(\ell_i,\ell_j)$ is required to be well separated 
%in the rapidity--azimuthal-angle plane, 
%\begin{equation}
%\label{eq:inc-cuts2}
% .
%\end{equation}
Photons from real Bremsstrahlung are recombined with the closest charged lepton 
if their separation obeys $\Delta R_{\ell_i,\gamma}< 0.2$.

\section{Phenomenological results}

\begin{table}
\begin{center}
\begin{tabular}
{|l|c|ccccc|}
\hline\vspace{-5mm}
&&&&&&\\
&$\sigma_{\bar q q}^\mr{LO}$~[fb] &  $\delta^\mr{weak}_{\bar q q}(\%)$&  
$\delta_{\bar q q}^\mr{phot}(\%)$&  $\delta_{\gamma\gamma}(\%)$ &$\delta_{q 
\gamma}(\%)$ & 
\\
\hline
\hline
%\\
% incl. [2$\mu$2e] & 11.4962(4) & $-4.32$ & $-0.93$ & $+0.1284$  & $+0.017$ & 
 [2$\mu$2e] & 11.4962(4) & $-4.32$ & $-0.93$ & $+0.13$  & $+0.02$ & 
\\
\hline
% incl. [4$\mu$]  &  \phantom{1}5.7308(3) & $-4.32$ & $-0.94$ & $+0.1145$ & $+0.019$ & 
[4$\mu$] &  \phantom{1}5.7308(3) & $-4.32$ & $-0.94$ &  $+0.11$ & $+0.02$ &
\\
\hline
\end{tabular}
\end{center}
\caption{LO cross sections for $\ppmmee$ and $\ppmmmm$ with the relative 
corrections $\delta_i =\sigma_i/\sigma_{\bar q q}^\mr{LO}$ 
for the LHC at $\sqrt{s}=13\TeV$.}
\label{tab:inc-xsec}
\end{table}

The results for the fiducial cross section in the setup described in the 
previous section are presented in Tab.~\ref{tab:inc-xsec} both for the 
mixed-flavour [2$\mu$2e] and equal-flavour [4$\mu$] final state. 
The first column gives the absolute prediction at LO. The equal-flavour case is 
approximately a factor of two smaller than the mixed-flavour case. Since 
the selection cuts described above are symmetric under exchange of the 
final-state leptons, the deviation from a naive symmetry factor of 
two can be directly attributed to interference terms that are only present 
in the equal-flavour case. From the ratio $\sigma_{\rm
  LO}[2\mu2\Pe]/(2\sigma_{\rm LO}[4\mu])\approx 1.003$ we find a
negative interference of about 0.3\% for the fiducial LO cross
section.

The four columns on the right-hand side of
Tab.~\ref{tab:inc-xsec} state the relative NLO EW corrections to the ${\bar q 
q}$ contribution and
the relative contributions of the photon-induced channels
$\delta_{\gamma\gamma}=\sigma_{\gamma\gamma}/\sigma_{\bar q q}^\mr{LO}$ 
and
$\delta_{q\gamma}=\sigma_{q\gamma}/\sigma_{\bar q q}^\mr{LO}$.
The ${\bar q q}$ contribution $\delta^\mr{EW}_{\bar q
  q}=\Delta\sigma_{\bar q q}^\mr{EW}/\sigma_{\bar q q}^\mr{LO}$ is
split into the purely weak and the photonic part,
$\delta^\mr{weak}_{\bar q q}$ and $\delta_{\bar q q}^\mr{phot}$,
respectively, so that $\delta^\mr{EW}_{\bar q
  q}=\delta^\mr{weak}_{\bar q q}+\delta_{\bar q q}^\mr{phot}$.
The relative corrections are at the level of the 
fiducial cross section equal for the mixed-flavour and the equal-flavour final 
state up to corrections at the sub-permille level. 
While this holds in the rather inclusive setup 
described in the previous section, this is not necessarily the case in other
phase-space
scenarios as has been pointed out in Ref.~\cite{Biedermann:2016lvg} for a setup 
with asymmetric selection cuts and a fiducial volume that enhances off-shell 
effects.

The bulk of the NLO corrections stems from the $\bar qq$ contribution with 
$-5.3\%$ which is dominated by the 
purley weak contribution. The impact of the purely photon-induced contribution 
$\delta_{\gamma\gamma}$ matters only at permille level. The contribution is  
suppressed via the small photon flux in the proton. Furthermore, in 
contrast to the $\bar q
q$-channel, there are no diagrams with two resonant Z-boson propagators present which 
additionally suppresses the $\gamma\gamma$ channel. The $q\gamma$-channel is even further 
suppressed by one order of magnitude and entirely negligible.

\begin{figure}
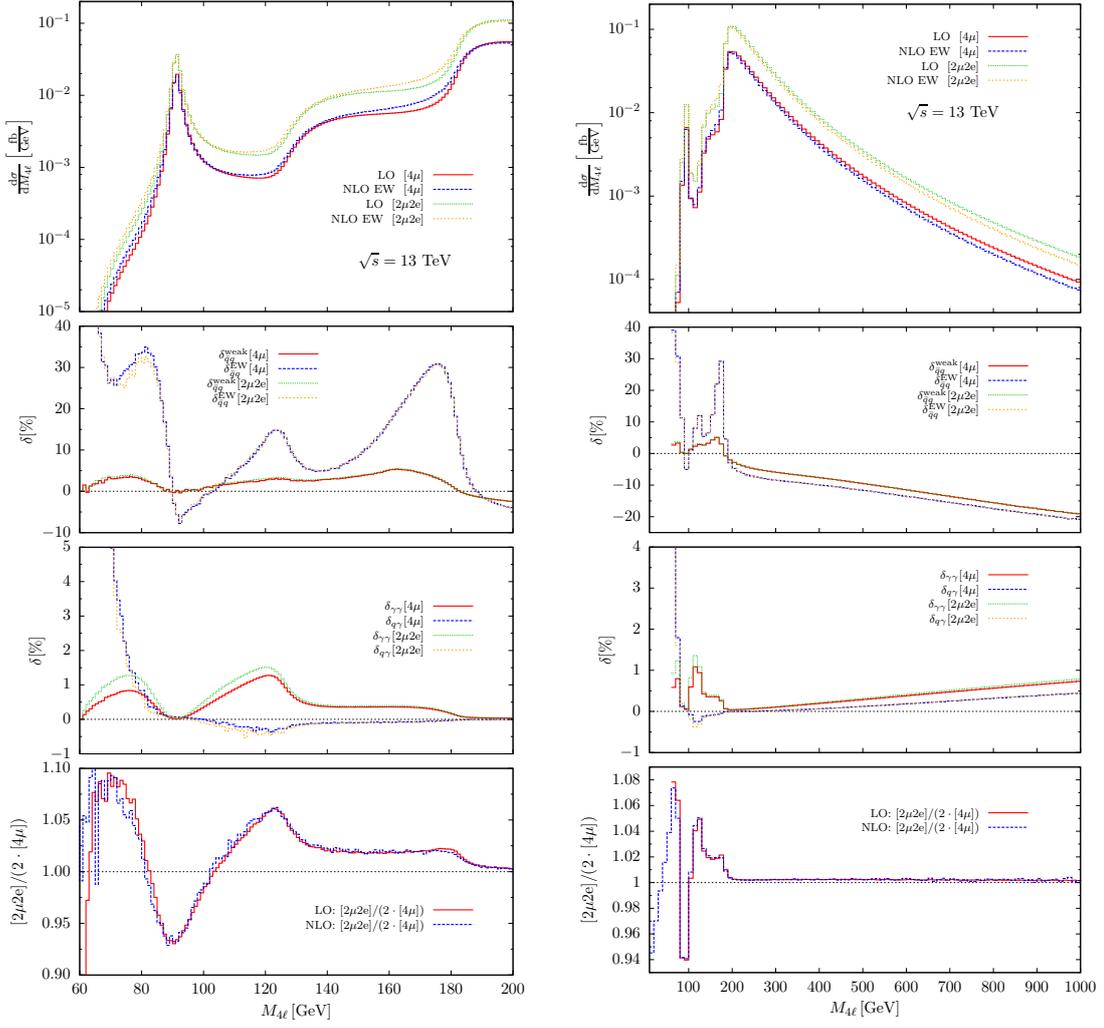

\begin{center}
\begin{minipage}{0.49\textwidth}
 
\includegraphics[width=\textwidth]{plots/{{ATLASincl.13TeV.invmass.4lep.ratio.4mu.2mu2e.highres.collsafe}}}
\end{minipage}
\begin{minipage}{0.49\textwidth}
  
\includegraphics[width=\textwidth]{plots/{{ATLASincl.13TeV.invmass.4lep.ratio.4mu.2mu2e.lowres.collsafe}}}
\end{minipage}
\end{center}
\caption{Invariant-mass distribution of the four-lepton system (upper panels),
   corresponding EW corrections (2nd panels from above), $\gamma\gamma$ and 
$q\gamma$
   contributions (third panels from above) for the unequal-flavour
   $[2\mu2\Pe]$ and the equal-flavour $[4\mu]$ final states in the
   inclusive setup. The panels at the bottom show the ratio of the $[2\mu2\Pe]$
   and $[4\mu]$ final states.}
\label{fig:inc-m4lep}
\end{figure}
As an example for a differential cross section, the invariant-mass distribution 
of the 
four-lepton system is shown in Fig.~\ref{fig:inc-m4lep}. The left column focuses on the 
off-shell sensitive region 
with its resonances and thresholds while the right column illustrates the whole 
spectrum up to 1~TeV. The absolute prediction in the upper row follows the 
chacteristic pattern of ZZ production: The first peak at $M_{4\ell}=M_\PZ$ 
represents the single Z-boson resonance followed by a decay into 
four charged leptons ($s$-channel configuration). The maximum at 
$M_{4\ell}=2M_\PZ$ represents the pair production threshold where both Z bosons 
may be produced on shell ($t$-channel configuration).
The knee 
above $M_{4\ell}=M_\PZ+2p_{{\rm T},{\rm min}}\approx120\GeV$ is
induced by the kinematical cut of Eq.~(\ref{eq:inc-cuts1}) on the lepton
transverse momentum. 

The panels below the absolute prediction show the relative EW corrections of the $\bar q
q$ channel and, separately, also the purely weak contribution. The EW 
corrections, or, more precisely, the photonic corrections,
exhibit the typical radiative tails in 
one-to-one correspondence to the resonance and threshold structure described 
above: Final state radiation of the real Bremsstrahlung photon may shift the 
value of $M_{4\ell}$ to lower values and thus shift the location of the 
resonance enhancement. Since the LO cross section falls off steeply below a 
resonance or a threshold, the relative correction becomes large and positive, 
and can reach several tens of percent. Above the ZZ production threshold, the 
photonic corrections are almost constant with $-2\%$ to $-3\%$. In contrast, 
the 
purely weak corrections strongly increase in magnitude up to $-20\%$ at 1\,TeV. This 
enhancement is due to electroweak logarithms that become large at high scales.
% 
% Although the 
% correction is due to electroweak logarithms, this is for this particular 
% observable not the Sudakov dominated region that requires all kinematic 
% invariants to be large. At high energies, however, ZZ production is dominated 
% by 
% Z bosons in forward/backwards direction where the corresponding kinematic 
% invariants are small. (An example for the Sudakov enhancement in ZZ production 
% with more than $-40\%$ corrections would be the lepton transverse momentum at 
% high scales, c.f. Fig.~9 in Ref.~\cite{Biedermann:2016lvg}). 
The purely weak corrections 
become positive below the pair production threshold and stay at the level of $0\%$ to $+5\%$ 
in the off-shell region. This non-trivial sign change is the most remarkable 
feature of the invariant-mass distribution and illustrates the need for the 
full 
off-shell computation of ZZ production: Because of this sign change it is impossible to include the NLO EW 
corrections via a global rescaling factor.

The third row of 
Fig.~\ref{fig:inc-m4lep} shows the impact of the photon-induced contributions. 
The $q\gamma$ contribution is strongly suppressed over the whole spectrum of 
the 
distribution, in agreement with the prediction for the fiducial cross section. 
Above the pair production 
threshold, the $\gamma\gamma$ channel contributes only at permille level. 
As the $\gamma\gamma$ contribution is a 
background contribution with at most one possibly resonant Z-boson propagator, the relative 
impact increases in the off-shell region up to the percent level. Since this region 
is dominated by the radiative tails of the photonic corrections, the impact of the 
$\gamma\gamma$ channel remains small. 

The panels in the lowest row display the size of the interfences at differential level
via the ratio $({\rm d}\sigma_{\rm (N)LO}[2\mu2\Pe]/{\rm
  d}M_{4\ell})/(2{\rm d}\sigma_{\rm (N)LO}[4\mu]/{\rm d}M_{4\ell})$. Above the pair-production
threshold at $M_{4
\ell}=2M_\PZ$, the interference is small and constant at the same level as for the 
fiducial cross section. In the region below $2M_\PZ$ where no lepton pair is 
resonant,
the size of the interference effect varies from $-7\%$ at $M_{4\ell}=M_\PZ$ to
$+6\%$ at $M_{4\ell}=M_\PZ+2p_{{\rm T},{\rm min}}$. In the region $M_\PZ+2p_{{\rm T},{\rm min}}\lesssim
M_{4\ell}\lesssim 2M_{\PZ}$, where only one lepton pair can be resonant,
the interference
effect amounts to $2\%$. Apart from minor effects at the sub-percent level, 
the NLO EW corrections do not affect the relative impact of the interferences.

More differential observables on ZZ production like transverse-momentum 
distributions, angular correlations, rapidity and invariant-mass 
distributions have been studied in detail in Ref.~\cite{Biedermann:2016lvg}, including 
also different event-selection criteria and the role of 
lepton pairing in the equal-flavour final state.

\section{Conclusion}

In this proceedings contribution we have presented selected results for the 
theoretical prediction of the production of four charged leptons at 
next-to-leading order accuracy in the electroweak coupling. The calculation 
includes all off-shell effects of intermediate virtual
photons and massive vector bosons. The purely weak corrections, a gauge invariant
subset of the full electroweak corrections,
may become large of the order of $-20\%$ to $-40\%$ in the high-energy limit 
due to logarithms of electroweak origin and stay at the level of $-5\%$ at intermediate scales.
The complementary photonic corrections give raise to large radiative tails 
near resonances or kinematical thresholds of several tens of percent. The photon-induced 
contributions turn out to be phenomenologically unimportant.
Comparing results of the mixed-flavour final state $\mu^+\mu^-\Pe^+\Pe^-$ with those for the equal-flavour
final state $\mu^+\mu^-\mu^+\mu^-$, we find interference effects in the off-shell sensitive region of the
order of $5\%$. The NLO EW corrections are an important ingredient for precision physics within the standard
model and its possible extensions, and need to be taken into account in future measurements.

\subsection*{Acknowledgements}

This work was supported by the German Federal Ministry for
Education and Research (BMBF) under contract no.~05H15WWCA1 and by the
German Science Foundation (DFG) under reference number DE 623/2-1.


\begin{thebibliography}{99}

\bibitem{Grazzini:2015hta}
M.~Grazzini, S.~Kallweit, and D.~Rathlev, {\it {ZZ production at the LHC:
  fiducial cross sections and distributions in NNLO QCD}},  {\em Phys. Lett.}
  {\bf B750} (2015) 407--410.
  %, [\href{http://arxiv.org/abs/1507.06257}{{\tt
  %arXiv:1507.06257}}].
  
  %\cite{Grazzini:2016swo}
\bibitem{Grazzini:2016swo}
  M.~Grazzini, S.~Kallweit, D.~Rathlev and M.~Wiesemann,
  {\it {$W^{\pm}Z$ production at hadron colliders in NNLO QCD}},
  Phys.\ Lett.\ B {\bf 761} (2016) 179.
  %  [arXiv:1604.08576 [hep-ph]].
  %%CITATION = doi:10.1016/j.physletb.2016.08.017;%%
  %46 citations counted in INSPIRE as of 28 Jun 2017
  
\bibitem{Grazzini:2016ctr}
  M.~Grazzini, S.~Kallweit, S.~Pozzorini, D.~Rathlev and M.~Wiesemann,
  {\it {$W^{+}W^{-}$ production at the LHC: fiducial cross sections and distributions in NNLO QCD}},
  JHEP {\bf 1608} (2016) 140.
%  [arXiv:1605.02716 [hep-ph]].
  %%CITATION = doi:10.1007/JHEP08(2016)140;%%
  %33 citations counted in INSPIRE as of 28 Jun 2017

\bibitem{Bierweiler:2013dja}
A.~Bierweiler, T.~Kasprzik, and J.~H. K{\"u}hn, {\it {Vector-boson pair
  production at the LHC to $\mathcal{O}(\alpha^3)$ accuracy}},  {\em JHEP} {\bf
  12} (2013) 071.
  %, [\href{http://arxiv.org/abs/1305.5402}{{\tt
  %arXiv:1305.5402}}].

\bibitem{Baglio:2013toa}
J.~Baglio, L.~D. Ninh, and M.~M. Weber, {\it {Massive gauge boson pair
  production at the LHC: a next-to-leading order story}},  {\em Phys. Rev.}
  {\bf D88} (2013) 113005.
  %, [\href{http://arxiv.org/abs/1307.4331}{{\tt
  %arXiv:1307.4331}}].

\bibitem{Biedermann:2016guo}
B.~Biedermann, et~al., {\it {Next-to-leading-order electroweak corrections to
  $pp \to W^+W^-\to4$~leptons at the LHC}},  {\em JHEP} {\bf 06} (2016) 065.
%  [\href{http://arxiv.org/abs/1605.03419}{{\tt arXiv:1605.03419}}].

\bibitem{Biedermann:2016yvs}
B.~Biedermann, A.~Denner, S.~Dittmaier, L.~Hofer, and B.~J\"ager, {\it
  {Electroweak corrections to $pp \to \mu^+\mu^-e^+e^- + X$ at the LHC: a Higgs
  background study}},  {\em Phys. Rev. Lett.} {\bf 116} (2016), no.~16 161803.
  %  [\href{http://arxiv.org/abs/1601.07787}{{\tt arXiv:1601.07787}}].
  
%\cite{Biedermann:2016lvg}
\bibitem{Biedermann:2016lvg}
  B.~Biedermann, A.~Denner, S.~Dittmaier, L.~Hofer and B.~Jager,
  {\it {Next-to-leading-order electroweak corrections to the production of four 
charged leptons at the LHC}},
  JHEP {\bf 1701} (2017) 033.
%  [arXiv:1611.05338 [hep-ph]].
  %%CITATION = doi:10.1007/JHEP01(2017)033;%%
  %6 citations counted in INSPIRE as of 28 Jun 2017

  \bibitem{Kallweit:2017khh}
  S.~Kallweit, J.~M.~Lindert, S.~Pozzorini and M.~Schonherr,
  {\it {NLO QCD+EW predictions for $2\ell2\nu$ diboson signatures at the LHC}},
  arXiv:1705.00598 [hep-ph].
  %%CITATION = ARXIV:1705.00598;%%
  %2 citations counted in INSPIRE as of 29 Jun 2017

%\bibitem{Aad:2015zqe}
%  G.~Aad {\it et al.} [ATLAS Collaboration],
%  {\it{Measurement of the $ZZ$ Production Cross Section in $pp$ Collisions at $\sqrt{s}$ = 13 TeV with the ATLAS Detector}},
%  Phys.\ Rev.\ Lett.\  {\bf 116} (2016) no.10,  101801.
%  [arXiv:1512.05314 [hep-ex]].
  
  \bibitem{Catani:1996vz}
S.~Catani and M.~Seymour, {\it {A general algorithm for calculating jet
  cross-sections in NLO QCD}},  {\em Nucl.Phys.} {\bf B485} (1997) 291--419.
%  [\href{http://arxiv.org/abs/hep-ph/9605323}{{\tt hep-ph/9605323}}].

\bibitem{Dittmaier:1999mb}
S.~Dittmaier, {\it {A general approach to photon radiation off fermions}},
  {\em Nucl.Phys.} {\bf B565} (2000) 69--122.
%  [\href{http://arxiv.org/abs/hep-ph/9904440}{{\tt hep-ph/9904440}}].

%\cite{Actis:2016mpe}
\bibitem{Actis:2016mpe}
  S.~Actis, A.~Denner, L.~Hofer, J.~N.~Lang, A.~Scharf and S.~Uccirati,
  {\it{RECOLA: REcursive Computation of One-Loop Amplitudes}},
  Comput.\ Phys.\ Commun.\  {\bf 214} (2017) 140.
%  doi:10.1016/j.cpc.2017.01.004
%  [arXiv:1605.01090 [hep-ph]].
  %%CITATION = doi:10.1016/j.cpc.2017.01.004;%%
  %20 citations counted in INSPIRE as of 29 Jun 2017

\bibitem{Denner:2016kdg}
  A.~Denner, S.~Dittmaier and L.~Hofer,
  {\it{Collier: a fortran-based Complex One-Loop LIbrary in Extended Regularizations}},
  Comput.\ Phys.\ Commun.\  {\bf 212} (2017) 220.
%  [arXiv:1604.06792 [hep-ph]].
  %%CITATION = doi:10.1016/j.cpc.2016.10.013;%%
  %46 citations counted in INSPIRE as of 29 Jun 2017
   
\bibitem{Accomando:2005ra}
E.~Accomando, A.~Denner, and C.~Meier, {\it {Electroweak corrections to $W
  \gamma$ and $Z \gamma$ production at the LHC}},  {\em Eur. Phys. J.} {\bf
  C47} (2006) 125--146.
  %[\href{http://arxiv.org/abs/hep-ph/0509234}{{\tt
  %hep-ph/0509234}}].

  
\bibitem{Denner:2005fg}
A.~Denner, S.~Dittmaier, M.~Roth, and L.~Wieders, {\it {Electroweak corrections
  to charged-current $e^+ e^- \to 4$ fermion processes: Technical details and
  further results}},  {\em Nucl.Phys.} {\bf B724} (2005) 247--294.
  
\bibitem{Ball:2013hta}
{\bf NNPDF} Collaboration, R.~D. Ball, et~al., {\it {Parton distributions with
  QED corrections}},  {\em Nucl. Phys.} {\bf B877} (2013) 290--320.
%  [\href{http://arxiv.org/abs/1308.0598}{{\tt arXiv:1308.0598}}].
 

\end{thebibliography}
\end{document}